# Dynamic exchange via spin currents in acoustic and optical modes of ferromagnetic resonance in spin-valve structures

A.A. Timopheev[1], Yu.G. Pogorelov[2], S. Cardoso[3], P.P. Freitas[3], G.N. Kakazei[2,4], N.A. Sobolev[1]

[1]*Departamento de Física and I3N, Universidade de Aveiro, 3810-193 Aveiro, Portugal*
[2]*IFIMUP and IN-Institute of Nanoscience and Nanotechnology, Departamento de Física e Astronomia, Universidade do Porto, 4169-007 Porto, Portugal*
[3]*INESC-MN and IN-Institute of Nanoscience and Nanotechnology, 1000-029 Lisbon, Portugal*
[4]*Institute of Magnetism, NAS of Ukraine, 03142 Kiev, Ukraine*

e-mail: andreyt@ua.pt

Two ferromagnetic layers magnetically decoupled by a thick normal metal spacer layer can be, nevertheless, dynamically coupled via spin currents emitted by the spin-pump and absorbed through the spin-torque effects at the neighboring interfaces. A decrease of damping in both layers due to a partial compensation of the angular momentum leakage in each layer was previously observed at the coincidence of the two ferromagnetic resonances. In case of non-zero magnetic coupling, such a dynamic exchange will depend on the mutual precession of the magnetic moments in the layers. A difference in the linewidth of the resonance peaks is expected for the acoustic and optical regimes of precession. However, the interlayer coupling hybridizes the resonance responses of the layers and therefore can also change their linewidths. The interplay between the two mechanisms has never been considered before. In the present work, the joint influence of the hybridization and non-local damping on the linewidth has been studied in weakly coupled NiFe/CoFe/Cu/CoFe/MnIr spin-valve multilayers. It has been found that the dynamic exchange by spin currents is different in the optical and acoustic modes, and this difference is dependent on the interlayer coupling strength. In contrast to the acoustic precession mode, the dynamic exchange in the optical mode works as an additional damping source. A simulation in the framework of the Landau-Lifshitz-Gilbert formalism for two ferromagnetic layers coupled magnetically and by spin currents has been done to separate the effects of the non-local damping from the resonance modes hybridization. In our samples both mechanisms bring about linewidth changes of the same order of magnitude, but lead to a distinctly different angular behavior. The obtained results are relevant for a broad class of coupled magnetic multilayers with ballistic regime of the spin transport.

1. Introduction

Spin current, a flow of angular momentum, is a basic concept in spintronics and spin caloritronics [1, 2]. Spin current generation is experimentally accessible via spin pumping [3-5], spin Seebek effect [6], spin Hall effect [7, 8] and acoustic wave propagation in the case of magnetic insulators [9]. The spin-orbit interaction plays a fundamental role in these effects. The presence of a spin current in a normal metal (NM) or semiconductor can be detected by the inverse spin Hall effect [10-12] or as a change of the effective damping in an adjacent ferromagnetic (FM) layer [3-5]. The latter effect allows one to alter the switching field of the FM layer and even sustain a stable precession in it [13-15]. It is hard to overestimate the fundamental and practical importance of the issues emerging from the investigation of the spin currents.

A precessing magnetic moment in a FM layer acts as a spin battery [16] injecting a pure spin current in a neighboring NM layer through the FM/NM interface. This spin current can then return to the NM/FM interface bringing the carried angular momentum back to the precessing spins of the FM layer. Depending on the spin-orbit interaction strength and the layer thickness, the normal metal will absorb a certain part of the angular momentum flow via the spin-flip relaxation processes. Thus, the backflow through the NM/FM interface will be always weaker than the direct flow, which results in an enhanced precession damping [3-5]. The spin diffusion length of the normal metal and the spin mixing interface conductance can be evaluated in this way [3-5, 17].

An interesting result has been obtained for a FM/NM/FM trilayer [18] having non-identical FM layers. The asymmetry provided different angular dependences of the ferromagnetic resonance (FMR) fields of the FM layers. When the external magnetic field was directed at an angle for which the FMR peak positions coincide, a narrowing of both resonances was observed. The explanation of this effect is that, for the case of separately precessing FM layers, the spin current generated in a precessing FM layer is absorbed in the other, non-resonating FM layer, which causes, in a full analogy to the written above, a damping enhancement, while for the case of a mutual resonant precession this spin current leakage is partially compensated by the spin current from the other FM layer. In this experiment, the NM spacer was thin enough for the spin current to be considerable at the second NM/FM interface, but thick enough to exclude any possible magnetic coupling between the FM layers.

Indeed, the magnetic coupling between two FM layers complicates the analysis of the spin-current-induced non-local damping. If the coupling is strong enough, the resonance response of the system is represented by the collective acoustic and optical modes which are the in-phase and out-of-phase mutual precession modes in the FM layers. There is no separate precession in such a regime – the precession in one layer drags the magnetic moment in the other one. Moreover, the linewidths of the resonance peaks are dependent on the field separation between them, and usually these parameters are angular dependent. And finally, the interaction fundamentally forbids the peaks to have a crossing point, i.e. the anticrossing is a characteristic feature here. The stronger the interlayer coupling, the larger is the anticrossing separation between the modes. From this point of view, the difference of damping for the acoustic and optical modes in a FM/NM/FM trilayer as a result of a dynamic spin currents exchange, theoretically predicted by Kim and Chappert [19], seems to be experimentally unachievable. Nevertheless, in several recent papers [20-22] experimental observations of this effect have been already claimed. There is, however, a full ignorance of the fact that the FMR peaks hybridization will also influence the linewidth even if a separate measurement of the precession in each layer can be done.

Motivated by this, we have performed a comprehensive study of weakly coupled spin-valve (SV) multilayers, where the hybridization is weak and the layers behave almost independently, conserving at the same time the main features of the acoustic and optical modes of the collective magnetic response. One important objective is to separate the hybridization-induced change of the FMR linewidth from the spin-current-induced one and to check in this way the difference between the spin-current-induced damping in the optical and acoustic regimes of precession. We present an experimental study of the FMR in NiFe/CoFe/Cu/CoFe/MnIr SV multilayers conducted using a standard X-band EPR spectrometer. Our study is accompanied by a simulation of the microwave absorption in such a magnetically coupled system in the presence of dynamical exchange by spin currents in the framework of the Landau-Lifshitz-Gilbert formalism.

## 2. Experimental details

FMR was measured at room temperature using a Bruker ESP 300E ESR spectrometer at a microwave frequency of 9.67 GHz. The first derivative of the microwave absorption by the magnetic field was registered. For each sample, a series of in-plane FMR spectra were collected for different angles of the magnetic field in the film plane with respect to the internal exchange bias field. Each FMR spectrum, experimentally measured or simulated, was fitted by Lorentzian functions to obtain angular dependences of the resonance field and linewidth. The least-squares method was employed.

The SV multilayers were grown by the ion-beam deposition in a Nordiko 3000 system. The cobalt-iron fixed layer is exchange coupled to the MnIr antiferromagnet (AF), the free layer is a bilayer composed of a permalloy and a cobalt-iron sublayers, and the copper spacer separates the free and fixed layers. Two series of samples were used in the study:

1) Glass / Ta(30 Å) / $Ni_{80}Fe_{20}$(30 Å) / $Co_{80}Fe_{20}$(25 Å) / **Cu ($d_{Cu}$)** / $Co_{80}Fe_{20}$(25 Å) / $Mn_{82}Ir_{18}$(80 Å) / Ta(30 Å) – the average thickness of the copper spacer, $d_{Cu}$, varies from 17 to 28 Å in 1 Å steps.
2) Glass / Ta(30 Å) / **$Ni_{80}Fe_{20}$(56 – $d_F$) / $Co_{80}Fe_{20}$($d_F$)** / Cu(22 Å) / $Co_{80}Fe_{20}$(25 Å) / $Mn_{82}Ir_{18}$(80 Å) / Ta(50 Å) – the relative thicknesses of the permalloy and cobal-iron sublayers vary within the 56 Å thick free layer by setting the parameter $d_F$ to 8, 16, 24, 32 and 40 Å.

Additionally, separate free layers (Glass / Ta(30 Å) / **$Ni_{80}Fe_{20}$(56 – $d_F$) / $Co_{80}Fe_{20}$($d_F$)** / Cu(22 Å) / Ta(50 Å)) of the first and second series were grown to serve as reference samples.

The first series was already studied in Refs. [23, 24]. It has been shown that the samples with $t_{Cu}$ > 16 Å are in the weak coupling regime, and the main interlayer coupling mechanism here is Néel's "orange-peel" magnetostatic interaction [25]. When the copper spacer thickness grows from 17 to 28 Å, the interlayer coupling energy is reduced from $1.1 \times 10^{-2}$ erg/cm$^2$ to $4 \times 10^{-3}$ erg/cm$^2$, which corresponds to a variation of the effective interaction field on the free layer from 17 to 6 Oe.

The second series has a fixed metallic spacer thickness, $t_{Cu} = 22$ Å, while the free layer effective magnetization, $4\pi M_{eff}$, determined by the Kittel formula, gradually varies from 15 kG to 8.5 kG. In this way the angular dependence of the free layer resonance field can be vertically shifted with respect to that of the fixed layer.

### 3. Simulation of the microwave absorption spectrum

A SV is considered as a system of two coupled FM layers consisting of a free and a fixed layer with the thicknesses $d_1$, $d_2$, volume saturation magnetizations $M_{s1}$, $M_{s2}$, and in-plane uniaxial magnetic anisotropy constants $K_1$, $K_2$, respectively. The exchange coupling of the fixed layer to the AF layer with the interface coupling energy $E_{ex}$ is defined by a unidirectional anisotropy with the effective field $E_{ex}/(d_2 M_{s2})$. The easy axes of all three anisotropies lay in the sample plane and have the same direction along the magnetic field applied at annealing. The magnetizations in both layers are assumed to be uniform, thus the bilayer magnetic state is completely described by the unit vectors $\hat{\mathbf{m}}_1$, $\hat{\mathbf{m}}_2$ of their instantaneous directions. The layers are coupled by the Heisenberg exchange interaction, $E_{ic}$.

Then the magnetic energy density per unit area of the considered system can be written as:

$$U_{tot} = d_1 \left( 2\pi M_{s1} (\hat{\mathbf{m}}_1 \cdot \hat{\mathbf{n}})^2 - K_1 (\hat{\mathbf{m}}_1 \cdot \hat{\mathbf{u}})^2 - H_{ext} M_{s1} (\hat{\mathbf{m}}_1 \cdot \hat{\mathbf{h}}_0) - h_{mw} M_{s1} (\hat{\mathbf{m}}_1 \cdot \hat{\mathbf{h}}_1) \right) + \\ + d_2 \begin{pmatrix} 2\pi M_{s2} (\hat{\mathbf{m}}_2 \cdot \hat{\mathbf{n}})^2 - K_2 (\hat{\mathbf{m}}_2 \cdot \hat{\mathbf{u}})^2 - H_{ext} M_{s2} (\hat{\mathbf{m}}_2 \cdot \hat{\mathbf{h}}_0) - \\ -h_{mw} M_{s2} (\hat{\mathbf{m}}_2 \cdot \hat{\mathbf{h}}_2) \ -\dfrac{E_{ex}}{d_2 M_{s2}} (\hat{\mathbf{m}}_2 \cdot \hat{\mathbf{u}}) \end{pmatrix} - E_{ic} (\hat{\mathbf{m}}_1 \cdot \hat{\mathbf{m}}_2). \quad (1)$$

There are also included four unit vectors determining the spatial orientation of the effective fields: the easy axis $\hat{\mathbf{u}} \perp \hat{\mathbf{n}}$ of the uniaxial and unidirectional anisotropies (here $\hat{\mathbf{n}}$ is the normal to the multilayer plane), the direction $\hat{\mathbf{h}}_0$ of the external magnetic field $H_{ext}$, and the direction $\hat{\mathbf{h}}_1$ of the microwave magnetic field $h_{mw}$.

The spin-pump / spin-sink mechanism in our SVs is considered as follows. The CoFe/Cu and Cu/CoFe interfaces are assumed to be identical and to give rise to an effective spin mixing conductance in the FM1/NM/FM2 structure characterized by the parameter $A_{FNF}^{\uparrow\downarrow}$ [26] which is in a generic case dependent on the relative magnetization orientations in the layers, $\hat{\mathbf{m}}_1$, $\hat{\mathbf{m}}_2$. Since the copper spacer is much thinner than the the spin-diffusion length ($\lambda_{sd} \sim 0.4$ µm at $T = 300$ K), the transfer of the angular momentum from one FM layer to the other occurs in a purely ballistic regime, i.e. the spin current emitted at the first CoFe/Cu interface is fully absorbed at the second Cu/CoFe interface. The spin current backflow is not considered separately: it just renormalizes the parameter $A_{\text{FNF}}^{\uparrow\downarrow}$. The spin-pump / spin-torque induced damping $\alpha_{\text{sp}}$ for each layer is influenced by its thickness, saturation magnetization and $g$-factor. The dynamics of such a structure can be described by a system of coupled Landau-Lifshitz-Gilbert equations with additional spin-pump / spin-torque induced Gilbert-like damping terms [5]:

$$\frac{\partial \hat{\mathbf{m}}_i}{\partial t} = -\gamma_1 \left[ \hat{\mathbf{m}}_i \times \mathbf{H}_{\text{eff}_i} \right] + \alpha_i \left[ \hat{\mathbf{m}}_i \times \frac{\partial \hat{\mathbf{m}}_i}{\partial t} \right] + \alpha_{\text{sp}_i} \left[ \hat{\mathbf{m}}_i \times \frac{\partial \hat{\mathbf{m}}_i}{\partial t} - \hat{\mathbf{m}}_j \times \frac{\partial \hat{\mathbf{m}}_j}{\partial t} \right],$$

$$\mathbf{H}_{\text{eff}_i} = \frac{1}{d_i M_{s_i}} \frac{\partial U_{\text{tot}}}{\partial \hat{\mathbf{m}}_i},$$

$$\alpha_{\text{sp}_i} = \mu_B g_i \frac{A_{\text{FNF}}^{\uparrow\downarrow}}{d_i 4\pi M_{s_i}},$$

$$i, j = 1, 2,$$

$$i \neq j.$$

(2)

The microwave field, $\propto h_{\text{mw}} e^{j\omega t} \hat{\mathbf{h}}_1$, is linearly polarized and directed along the multilayer normal, $\hat{\mathbf{h}}_1 \parallel \hat{\mathbf{n}}$, while the external static magnetic field lies in the film plane, $\hat{\mathbf{h}}_0 \perp \hat{\mathbf{n}}$, making an angle $\varphi_h$ with the system's easy axis $\hat{\mathbf{u}}$. A linear response of the system relates to small angle deviations from the equilibrium, $\hat{\mathbf{m}}_1 + \boldsymbol{\delta}\mathbf{m}_1$, $\hat{\mathbf{m}}_2 + \boldsymbol{\delta}\mathbf{m}_2$, $\boldsymbol{\delta}\mathbf{m}_1 \perp \hat{\mathbf{m}}_1$, $\boldsymbol{\delta}\mathbf{m}_2 \perp \hat{\mathbf{m}}_2$. The complex vectors $\boldsymbol{\delta}\mathbf{m}_1$, $\boldsymbol{\delta}\mathbf{m}_2 \propto e^{j\omega t}$ can be found from a linear 4×4 system by Eqs. (2) linearized near the equilibrium. This system is too complicated for an analytical treatment but easily solved numerically using a standard desktop computer. A certain simplification can be achieved using spherical coordinates. The microwave absorption is proportional to the imaginary part of the microwave susceptibility in the direction of the microwave field:

$$\chi'' = \frac{d_1 + d_2}{d_1 + d_2 + d_{\text{Cu}}} \text{Im} \left[ \frac{d_1 M_{s1} \left( \boldsymbol{\delta}\mathbf{m}_1 \cdot \hat{\mathbf{h}}_1 \right) + d_2 M_{s2} \left( \boldsymbol{\delta}\mathbf{m}_2 \cdot \hat{\mathbf{h}}_1 \right)}{h_{\text{mw}}(d_1 + d_2)} \right].$$

(3)

To treat the volume microwave susceptibility of a SV, the metallic spacer width, $d_{\text{Cu}}$, was added in Eq. (3). Then a full cycle of calculations in each simulation consists of: *i*) finding the equilibrium orientation of the magnetic moments by the minimization of Eq. (1); *ii*) numerical solution of Eq. (2) linearized near the equilibrium; *iii*) combining the obtained precession amplitudes in the volume susceptibility by Eq. (3). The separate susceptibility of each layer can be obtained if the thickness of the other layer is set to zero at the last step of calculations. This can be useful in the analysis of experimental data obtained by the element-specific X-ray magnetic circular dichroism, time-resolved Kerr microscopy and other techniques allowing to separately measure the microwave responses of the layers [22, 27, 28].

The magnetic parameters in our simulations were set in accordance to the experiment. In the studied samples, the in-plane effective fields of the free and fixed layers are several times lower than the resonance field of the free layer ($H_{res}$ > 600 Oe), whose FMR linewidth will be the main discussion issue in the present paper. This implies that at the free layer's resonance conditions the magnetic field almost aligns both magnetic moments. Thus, the dynamic exchange via spin currents will be considered in the collinear regime, and the parameter $A_{FNF}^{\uparrow\downarrow}$ is assumed to be independent of the in-plane magnetic field orientation.

### 4. General features of the FMR in both SV series

The dynamics of two coupled FM layers can be described in terms of acoustic and optical modes, a hybridized response of the system to the exciting microwave field. These modes are the in-phase and out-of-phase mutual precession of the magnetic moments in the FM layers. The acoustic mode bears averaged magnetic parameters of the system, while the optical one gives information about the system's asymmetry. The interlayer coupling shifts the optical mode away from the acoustic one, therefore, the coupling strength can be determined if the other effective fields in the system are known. However, this is a strong coupling regime which has few similarities with the FMR of standard SV multilayers, including the samples used in this study, where the effective interlayer coupling does not exceed several tens of Oersted.

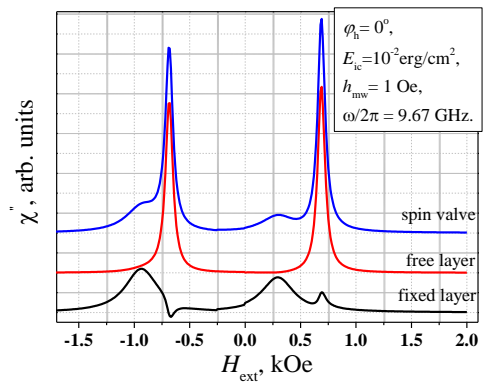

**Fig. 1.** FMR spectrum calculated for a SV in the weak coupling regime (top curve). The middle and bottom curves show separated responses from the free and fixed layers in the SV. The layer parameters correspond to the first series of SVs: $d_1$ = 5.5·nm, $\alpha_1$ = 0.012, $M_{s1}$ = 1155 emu/cm$^3$, $K_1$ = 5.7·10$^3$ erg/cm$^3$; $d_2$ = 2.5 nm, $\alpha_2$ = 0.055, $M_{s2}$ = 1175 emu/cm$^3$, $K_2$ = 1.7·10$^4$ erg/cm$^3$, $E_{ex}$ = 0.094 erg/cm$^2$ and $E_{ic}$ = 0.01 erg/cm$^2$.

The samples under study are in a weak coupling regime provided by Néel's "orange-peel" magnetostatic interaction [25]. The determined effective interlayer coupling field, acting from one layer to another, is in the 10 to 30 Oe range for both layers [24] in all samples of the two series. The main interaction effect is a constant decrease of the resonance field in each layer. This and other related effects are thoroughly discussed in Ref. [24].

To support the ideology of the weak coupling regime, a simulation of the microwave response has been done using a parameter set for the first series and the interlayer coupling strength $E_{ic} = 0.01$ erg/cm$^2$. The spin-pump / spin-sink mechanism was switched off: $\alpha_{sp1} = \alpha_{sp2} = 0$. Fig. 1 shows a typical microwave absorption spectrum of a SV multilayer and respective separated responses of each layer in it. The magnetic moments are precessing almost independently, and therefore each peak can be associated with the precession of the magnetization in a specific layer. The asymmetry of the thicknesses, damping parameters and magnetizations is clearly manifested in these spectra. A fixed layer with half the thickness of the free one is much easier dragged by the precessing free layer. However, the inverse effect, a drag of the free layer by the resonance precession in the fixed layer, is not so pronounced: only a small asymmetry on the wings of the free layer peak is observed. A four times stronger damping, mainly that due to the contact with an antiferromagnet [24], produces a much lower precession amplitude of the fixed layer. The situation gets even worse because the free layer is twice as thick as the fixed one, thus, the effective interlayer coupling field, acting on the free layer from the precessing fixed layer, is about two times lower. Leaping ahead, it is evident that the spin-pump / spin-torque effect will be more pronounced in the free layer.

A very important feature is that, despite the almost independent precession of the layers, an optical-like and acoustic-like behavior is still present in the dynamics. A precessing layer drags the magnetization of the other layer either in the "in-phase" or in the "out-of-phase" regime. For the case of ferromagnetically coupled layers, the optical mode (an out-of-phase mutual precession) has, in a given magnetic field, a higher precession frequency than the acoustic mode (an in-phase mutual precession). Therefore, the optical mode will be observed, at a given microwave frequency, in lower resonance fields. A specifics of the first sample series is that, for the parallel and antiparallel orientations of $H_{ext}$ ($\varphi_h = 0°$ and 180°), the resonance field of the fixed layer is respectively lower (~ 300 Oe) or higher (~ 1000 Oe) than that of the free layer (~ 700 Oe in both cases). As seen from Fig. 1, this brings about an interesting behavior: the precession of the free layer in the parallel $H_{ext}$ ($H_{ext} > 0$, $\varphi_h = 0°$) drags the fixed layer "in-phase", while in the antiparallel orientation ($H_{ext} < 0$, $\varphi_h = 0°$) it drags the fixed layer "out-of-phase", i.e. in the optical mode.

It is evident that the switching between the acoustic and optical "drag" regimes would disappear with the fixed layer resonance peak being below that of the free layer. This justifies our choice of the sample series: a variation of the interlayer coupling in the first series should influence the intensity of the dragged precession, while varying the effective magnetization of the free layer in the second series will tune the resonance field of the free layer with respect to that of the fixed one.

Fig. 2 shows the evolution of the angular dependences of the resonance field in both series. The general properties of the samples are as follows. The effective field of unidirectional anisotropy for the fixed layer is about 300 Oe, and it is the main in-plane anisotropic contribution here. The free layer has a weak in-plane unidirectional anisotropy of 5 to 20 Oe, varying with the NiFe/CoFe composition. The magnetic parameters of the free layer are less fluctuating than those of the fixed one since the former is thicker and always deposited on the same surface. The increased roughness of the fixed layer also strongly influences the AF/FM interface, giving rise to fluctuations not only of the fixed layer's effective magnetization but also of the exchange bias coupling. It is hard as well to prepare reference samples for the fixed layer. Our previous investigation has shown that a separately deposited fixed layer has considerably different magnetic parameters [24]. The strong angular variation of the resonance field and the direct contact with the AF has also a strong influence on the angular dependence of the linewidth even in a separately deposited fixed layer. Moreover, as the linewidth is extracted using the least-squares method, the accuracy of the fitting for the low-intensity peak stemming from the fixed layer will be much lower than for the free layer. Due to these reasons and the asymmetry discussed above, the following discussion of the experimental results is mostly focused on the linewidth, $\Delta H_{fr}$, of the free-layer-related peak and on its angular dependence, $\Delta H_{fr}(\varphi_h)$.

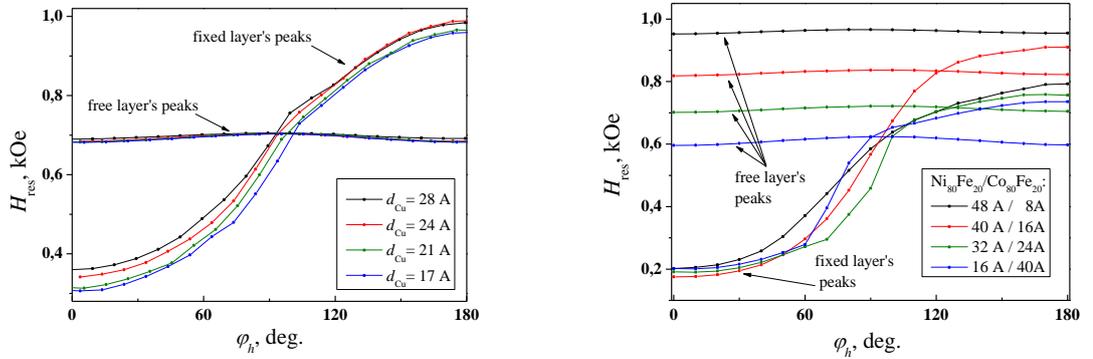

**Fig. 2.** Angular dependences of the FMR peaks for the free and fixed layer: the first series where the interlayer coupling strength is varied by gradually changing the metal spacer thickness $d_{Cu}$ (left panel); the second series where the mean FMR field of the free layer is varied by gradually changing the free layer effective magnetization, $M_{s1}$ (right panel).

## 5. Analysis of angular dependences

Additional reference samples which completely duplicate the free layer and the next nearest nonmagnetic layers in each SV sample have been grown and used as a reference in the analysis of the angular dependences of the free layer FMR linewidth, $\Delta H_{\text{fr}}(\varphi_h)$. It has been found that $<\Delta H_{\text{fr}}>$ (averaged over the whole $\varphi_h$ range) of each reference sample is at least 20% lower than $<\Delta H_{\text{fr}}>$ in the corresponding SV sample. However, the increased damping in the presence of a second FM layer (i.e. fixed layer) cannot be uniquely associated with the spin-pump / spin-sink mechanism [5, 26], because a non-zero interlayer coupling causes a hybridization of the resonance modes. Though the layers are weakly coupled, each layer's resonance mode bears a small portion of the magnetic behavior of the layer coupled to it. As the free layer's damping parameter is several times lower than the fixed-layer-related one, the observed FMR line broadening in the SV can have both origins, and it demands a quantitative analysis. At the same time, the shape of the $\Delta H_{\text{fr}}(\varphi_h)$ dependence in the SV samples deserves additional attention.

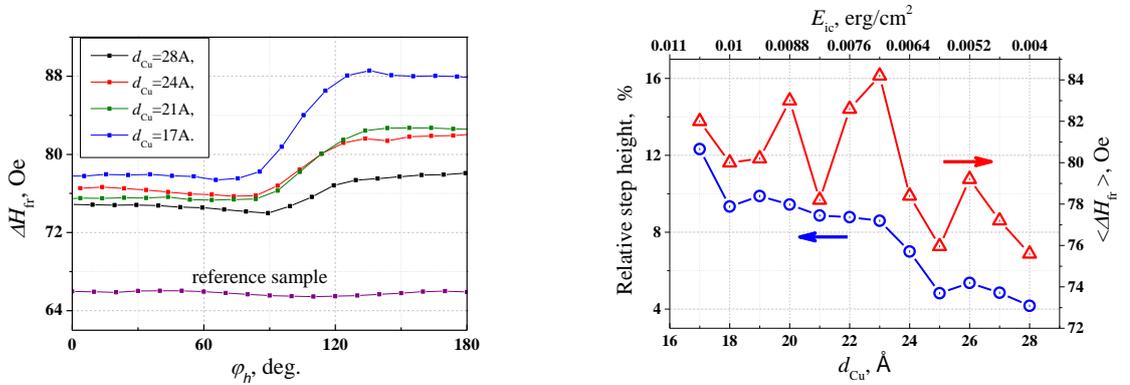

**Fig. 3.** Linewidth of the free layer in the first sample series. Left panel: The angular dependence for different copper spacer thicknesses. The reference sample curve does not show steps. Right panel: The relative step height and mean linewidth versus the interlayer coupling strength.

Fig. 3 shows experimental results obtained on the first series of samples, where the interlayer coupling has been gradually tuned by changing the copper spacer thickness. The reference layer does not show any noticeable $\Delta H_{\text{fr}}(\varphi_h)$ dependence. In contrast, a step-like shape of the $\Delta H_{\text{fr}}(\varphi_h)$ dependence has been observed in all SVs. A noticeable growth of $\Delta H_{\text{fr}}$ is observed for the antiparallel orientation of the magnetic field ($90° < \varphi_h < 270°$). The transition from the weaker damped to the stronger damped regime is quite smooth and occurs within the angular range where the fixed layer peak crosses the free layer's one (see Fig. 2). The relative step height in the $\Delta H_{\text{fr}}(\varphi_h)$ dependence has been found to decrease with increasing copper spacer thickness, $d_{\text{Cu}}$. In other words, with decreasing interlayer coupling, assumed to be the only parameter influencing the free layer in this series, the observed step height also decreases. As seen from Fig. 3, the relative step height monotonously decreases from 12% to 4% with decreasing interlayer coupling. It should be noted that, among the other extracted SV parameters analyzed as a function of $d_{Cu}$, this one has the smoothest dependence. As an example, we show the thickness dependence of $\langle \Delta H_{fr} \rangle$ averaged over the whole [0, 360°] range of angles (Fig. 3). Though the scattering of experimental points is several times higher, this parameter also shows a tendency to decrease, whose nature is hard to identify at present. A degree of resonance modes hybridization, weakening with decreasing interlayer coupling, seems to be the most probable source of this effect. The free layer resonance precession drags the magnetic moment of the fixed layer, and this could be itself an additional source of increased linewidth. A more detailed discussion of a simultaneous influence of hybridization and spin-pump / spin-sink effects on the linewidth will be given in the next Section.

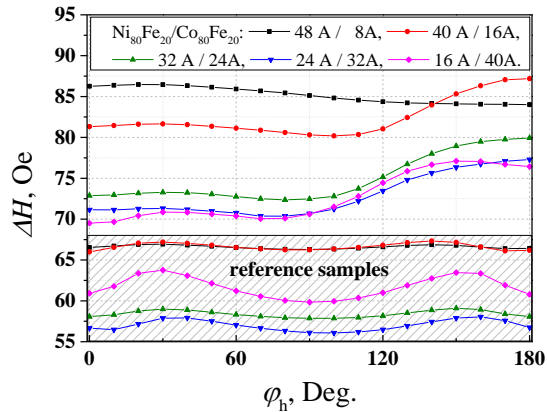

**Fig. 4.** Angular dependences of the linewidth for the free layer in the second sample series and in the respective reference samples.

In the second SV series, an increase of the $\langle \Delta H_{fr} \rangle$ parameter in comparison with the reference layers is also clearly seen (see Fig. 4). At the same time, the observed step-like $\Delta H_{fr}(\varphi_h)$ dependence has revealed additional features. The step from the weaker damped to stronger damped regime is shifted to higher angles as the mean resonance field of the free layer gets higher. The observed shift completely matches that of the crossing angle, i.e., the angle where the resonances of the free and fixed layers coincide (see Fig. 2). The most important feature is the absence of step-like behavior in the $\Delta H_{fr}(\varphi_h)$ dependence for the sample with the $Ni_{80}Fe_{20}$(48 Å) / $Co_{80}Fe_{20}$(8 Å) free layer. Fig. 2 shows that the resonances are not crossing there at all: the free layer's resonance field is always higher than the fixed layer's one.

As compared to the first series, there are also additional peculiarities in the $\Delta H_{fr}(\varphi_h)$ dependences, distorting the step-like shape. They, however, are linked to the intrinsic angular dependence of $\Delta H_{fr}$ of a concrete free layer. An analysis of the reference samples shows that the increase of the $Co_{80}Fe_{20}$ / $Ni_{80}Fe_{20}$ thickness ratio causes a noticeable increase in the angular variation of $\Delta H_{fr}$. Also a considerable variation of the damping parameter is observed in the reference samples, however, of a nonsystematic character. These intrinsic features, as seen from Fig. 4, are conserved also in the SV samples.

Thus, the observed experimental results can be resumed as follows. When the fixed layer resonance field is higher than the free layer's one, the linewidth of the free layer peak, $\Delta H_{\text{fr}}$, gets larger. The respective angular dependence, $\Delta H_{\text{fr}}(\varphi_h)$, shows a step-like shape with the threshold angular position corresponding to the crossing region of the free and fixed layer resonances. The step height decreases with decreasing interlayer coupling strength. This effect is absent in the reference samples containing only the free layer, as well as it disappears in the SVs where the resonances of the free and fixed layers do not cross.

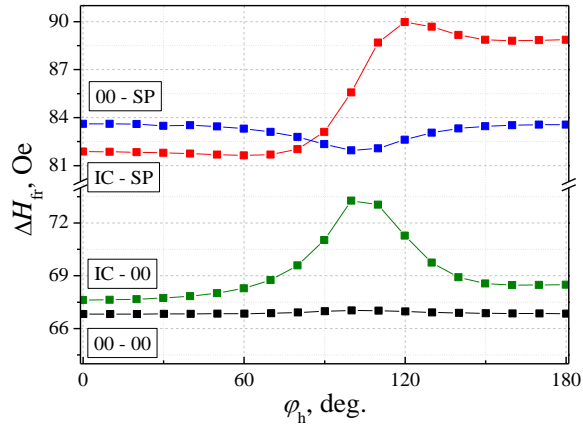

**Fig. 5.** Simulated angular dependences of the free layer's FMR linewidth. Four different regimes are shown: "00-00": $E_{\text{ic}} = 0$ and $A_{\text{FNF}}^{\uparrow\downarrow} = 0$; "IC-00": $E_{\text{ic}} = 0.01$ erg/cm$^2$ and $A_{\text{FNF}}^{\uparrow\downarrow} = 10$; "00-SP": $E_{\text{ic}} = 0$ and $A_{\text{FNF}}^{\uparrow\downarrow} = 1.1\times10^{15}$ cm$^{-2}$; "IC-SP": $E_{\text{ic}} = 0.01$ erg/cm$^2$ and $A_{\text{FNF}}^{\uparrow\downarrow} = 1.1\times10^{15}$ cm$^{-2}$. The layer parameters refer to the first series of SVs, as they are already listed in the caption to Fig. 1.

### 6. Hybridization versus non-local damping

To clarify the interpretation of the experiment, a series of in-plane FMR spectra were simulated as a function of the in-plane magnetic field direction $\varphi_h$ employing the formalism described in Sec. 3. The simulated spectra display the resonance peaks by the free and fixed layer (as shown, e.g., in Fig. 1). By fitting a set of overlapping Lorentzians to the simulated spectrum, the resonance peaks' parameters were deduced. Then the angular dependence of the linewidth of the free layer, $\Delta H_{\text{fr}}(\varphi_h)$, was analyzed. For the first sample series, the layer parameters and coupling were determined in our previous work [24] on exactly the same samples. For the second series, these parameters were chosen to reproduce the experiment as close as possible, and the interlayer coupling was fixed to $E_{\text{ic}} = 0.01$ erg/cm$^2$ in all SVs. Fluctuating parameters of the fixed layer and a slight variation of the internal damping of the free layer noted in the experiment were ignored in the simulation. In both series, the effective spin-mixing conductance for the whole FM/NM/FM structure is assumed to be $A_{FNF}^{\uparrow\downarrow} = 1.1\times10^{15}$ cm$^{-2}$ (which is slightly lower than in case of a single Co/Cu interface ~ $1.4\times10^{15}$ cm$^{-2}$ [26]), in units of $e^2/h$.

Relative contributions of the hybridization and spin-pump / spin-sink effects to the linewidth of a weakly coupled SV system are the central object of the present investigation. Referring to a SV from the first series, we have done four different simulations (see Fig. 5) of the $\Delta H_{fr}(\varphi_h)$ dependences. First, both the interlayer coupling (IC) and the spin mixing conductivity (SP) were set to zero (the "00-00" curve). This has demonstrated that the fitting procedure correctly extracts the linewidth, and the free layer's $\Delta H_{fr}$ does not depend on the peaks separation between the free and fixed layers (when fully uncoupled). It has been found that a small increase of $\Delta H_{fr}$ is observed in the crossing region. This increase, however, is lower than 0.3%, thus being at least one order of magnitude lower than the other factors relevant for the $\Delta H_{fr}(\varphi_h)$ dependence, both in the experiment and simulation. Therefore, this factor was ignored in the above experimental data and will be omitted in the further considerations.

The next simulation has been made with $E_{ic} = 0.01$ erg/cm$^2$ and $A_{FNF}^{\uparrow\downarrow} = 0$ (the "IC-00" curve). In this case, a noticeable increase (~ 7%) in $\Delta H_{fr}$ is observed in the crossing region. This effect can be only attributed to an enhanced hybridization of the resonance peaks in this region. When increasing the linewidth of the free layer peak, the hybridization also makes the fixed layer peak narrower. The dependence of the hybridization degree on the distance between the resonance peaks is also responsible for the fact that the $\Delta H_{fr}$ value for the antiparallel orientation ($\varphi_h = 180°$) is slightly higher (by ~ 1.3%) than that for the parallel orientation ($\varphi_h = 0°$). As seen from Fig. 2, the resonance peaks are indeed closer to each other in the antiparallel orientation. It is worth noting that the shape of the $\Delta H_{fr}(\varphi_h)$ dependence is quite different from the experimentally observed step-like profile.

A pure spin-pump / spin-sink regime has been set in the next simulation, i.e. with $E_{ic} = 0$ and $A_{FNF}^{\uparrow\downarrow} = 1.1\times10^{15}$ cm$^{-2}$. The corresponding $\Delta H_{fr}(\varphi_h)$ dependence is labeled "00-SP". In comparison with the previously discussed regime, $\Delta H_{fr}$ is depressed (by ~ 2%) in the crossing region. This effect was observed experimentally in a FM/NM/FM system and has been interpreted as a partial compensation of the spin current leakage which occurs when both FM layers are in resonance precession [5] and thus emit the spin currents. Without discussing this in details, we note only two points: *i)* due to the considerably thicker FM layers in our SVs, the observed effect is much weaker than in the above mentioned paper [5]. Since the spin torque effect is of interfacial origin, its influence scales with the inverse layer thickness; *ii)* the spin-pump / spin-sink and hybridization effects work in the opposite senses in the crossing region.

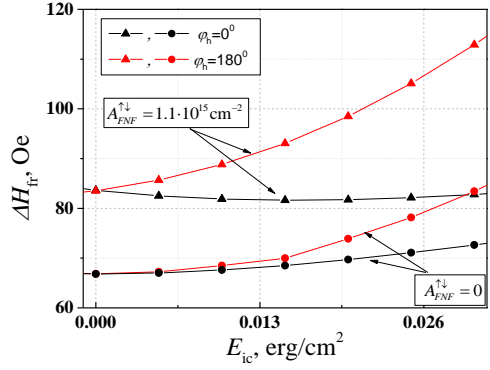

**Fig. 6.** Linewidth of the free layer in the parallel and antiparallel orientation versus the interlayer coupling strength simulated through spin conductivity (and without it). The layer parameters are set for the first series of SVs, as they are already listed in the caption to Fig. 1.

The last simulation, labeled "IC-SP", shows a simultaneous action of the interlayer coupling and spin-pump / spin-sink effect, i.e. $E_{ic}$ = 0.01 erg/cm$^2$ and $A_{FNF}^{\uparrow\downarrow}$ = 1.1×10$^{15}$ cm$^{-2}$. As seen from Fig. 5, there is a good agreement with the experiment. The step size in the $\Delta H_{fr}(\varphi_h)$ dependence is ~ 8%, also very close to the experimental values. In the parallel orientation ($\varphi_h$ = 0º), the $\Delta H_{fr}$ value is almost the same as in the crossing region for the case of the pure spin-pump / spin-sink effect. This means that a partial compensation of the spin current leakage takes place in the whole range of angles for the acoustic regime of precession (–90º < $\varphi_h$ < 90º). On the contrary, in the optical regime (–110º > $\varphi_h$ > 110º) the free layer suffers additional damping, absent in the previously discussed "00-SP" simulation. The explanation is as follows. The precession can be geometrically separated in a transversal and a longitudinal component of magnetization with respect to its equilibrium orientation. The conservation of angular momentum allows the same separation for the generated spin current. For a small-angle precession, the transversal component of magnetization ($\propto \sin(\theta_{prec})$) is larger than the longitudinal one ($\propto \sin^2(\theta_{prec}/2)$). The transversal part varies in time, while the longitudinal does not (at least in the linear response approximation, neglecting, e.g., a possible nutation). The importance of the time-dependent transversal part of the spin current has been recently shown in Ref. [29]. Both components are transferred by the spin current from one FM layer to the other. In the acoustic precession mode (as well as in the crossing point for the "00-SP" case), the transversal component of the spin current from the second layer is in-phase with the transversal part of that from the first layer. Therefore, the spin current absorbed at the interface should act in an "anti-damping" manner. On the contrary, in the optical precession regime the transversal component of the absorbed spin current is out-of-phase with the magnetic moment precession, and therefore an extra damping occurs. An increase of the non-local damping in the optical precession regime in a magnetically coupled FM/NM/FM trilayer has been predicted by Kim in Ref. [19]. Probably this effect was observed in several papers [20-22]. However, its interpretation in these papers fully ignores the hybridization of resonance modes, and therefore it is hard to draw some clear conclusions.

The weak interlayer coupling and an almost symmetrical position of the free layer peak with respect to the fixed one in the first SV series play an important role in the non-local damping effect. Fig. 6 shows the calculated $\Delta H_{fr}$ parameter versus the interlayer coupling strength for $\varphi_h = 0°$ and $\varphi_h = 180°$, with and without spin-pump / spin-sink effect. It is seen that, for $E_{ic} < 0.013$ erg/cm$^2$, the increase of $\Delta H_{fr}$ occurs merely due to the non-local damping effect, while for a stronger coupling the hybridization takes a comparable role, and these two contributions are hardly separable in a real experiment. From this simulation it is also seen that the dynamic exchange via spin currents is quite different in the optical and acoustic precession modes. The increase of $\Delta H_{fr}$ due to increasing hybridization is suppressed in the acoustic mode ($\varphi_h = 0°$) by "anti-damping", i.e., in-phase interaction between the transversal components of magnetization and the absorbed spin current. On the contrary, in the optical precession mode ($\varphi_h = 180°$) the effect of non-local damping is considerably enhanced, as the transversal components of the precessing magnetization and of the absorbed spin current are out-of-phase.

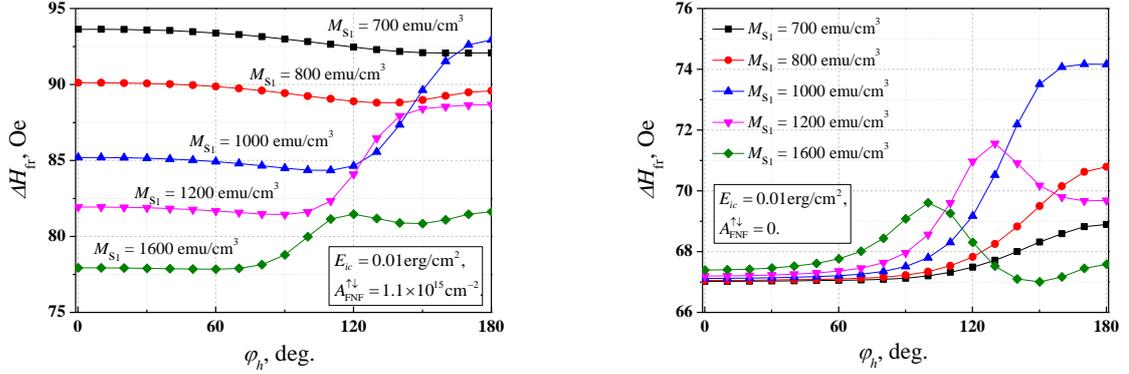

**Fig. 7.** Angular behavior of the linewidth in the second series of SVs, with a gradual variation of the effective magnetization of the free layer, simulated considering the spin conductivity and without it. For the red and black curves, the fixed layer resonance does not cross that of the free layer anymore. The parameters set is the same as for the first series and with $M_{s2} = 1525$ emu/cm$^3$ and $E_{ex} = 0.12$ erg/cm$^2$.

A simulation of the $\Delta H_{fr}(\varphi_h)$ dependence in the second series, where the effective magnetization of the free layer, $M_{s1}$, is gradually changed, completes the discussion. A comparison of the simulation (Fig. 7) with the experiment (Fig. 2, right panel) allows one to conclude that the effects of non-local damping are also clearly seen here. First, when the free layer's saturation magnetization is such low that the fixed layer peak does not cross the free layer resonance, and therefore, the precessing free layer drags the fixed layer always in-phase (acoustic mode), a characteristic step-like feature in the $\Delta H_{fr}(\varphi_h)$ dependence disappears. In these regime, the calculated $\Delta H_{fr}(\varphi_h)$ dependences are fundamentally different, irrespectively of whether the spin conductivity exists in the system or not. For the case of $A_{FNF}^{\uparrow\downarrow} = 0$, the fixed layer peak approaching the free layer one at $\varphi_h$ = 180° induces an enhanced hybridization, and $\Delta H_{\text{fr}}$ grows, while for $A_{\text{FNF}}^{\uparrow\downarrow}$ = 1.1×10$^{15}$ cm$^{-2}$ the enhanced hybridization is fully suppressed by the described above "anti-damping" feature of the acoustical mode of precession in the presence of spin conductivity. A decrease of $\Delta H_{\text{fr}}$ is observed when the fixed layer peak is approaching. The closer is the fixed layer resonance to the free layer one, the higher is the precession amplitude in the fixed layer, and thus the higher is the generated spin current. Therefore, a decrease of $\Delta H_{\text{fr}}$ is observed. Another distinct feature of the non-local damping is a continuous growth of the low-angle part of the $\Delta H_{\text{fr}}(\varphi_h)$ dependence (which corresponds to the acoustical precession mode) with decreasing $M_{s1}$. As $M_{s1}$ decreases, all effective fields arising from the interface, as well as the spin torque emerging from the absorbed spin current, will increase. For the case of zero spin conductivity, the low-angle part of the $\Delta H_{\text{fr}}(\varphi_h)$ dependence remains always the same. Both these features are clearly seen in the experiment (Fig. 2, right panel).

## 7. Conclusions

In-plane angular dependences of the free layer's FMR linewidth have been studied in two series of spin-valve multilayers, where the free and fixed layers are weakly coupled by Néel's "orange peel" magnetostatic interaction. In the first series, the interlayer coupling strength was varied by changing the metal spacer thickness, while in the second series the in-plane resonance field of the free layer was tuned by changing the Ni$_{80}$Fe$_{20}$/Co$_{80}$Fe$_{20}$ thickness ratio.

The main experimental results are as follows. The angular dependence of the linewidth of the free layer displays a characteristic step-like feature. When the resonance field of the fixed layer is higher than that of the free layer, the damping increases. The transition from the weakly damped to strongly damped regime occurs in the angular region of the peaks crossing. The reference samples, containing only a free layer and an adjacent nonmagnetic layer, do not show such a behavior. Similarly, no step is observed in the samples from the second series, where the fixed layer peak does not cross that of the free layer at all. The step size decreases with decreasing interlayer coupling strength.

A comparison with simulations has shown that the observed effect is due to the non-local damping effect. In the weakly coupled regime, the hybridization of the resonance peaks is low, and each peak can be attributed to the resonance precession of a particular layer. At the same time, due to a non-zero magnetic coupling, the resonant precession in one layer induces a small correlated precession ("drag") in the other one. Depending on the relative field position of the free layer resonance peak with respect to the fixed one, the fixed layer magnetic moment is "dragged" either in the acoustic-like ("in-phase" precession in both layers) or optical-like ("out-of-phase" mutual precession) regime. Therefore, varying the in-plane angle between the external magnetic field and the exchange bias field and changing in this way the relative peaks field position, one can switch between these two regimes. In case of ballistic regime of spin transport, additionally to the time-independent longitudinal component, the spin current generated by the dragged fixed layer has a time-varying transversal component, being "in-phase" or "out-of-phase" with the time-varying transversal component of the free layer's precessing magnetization. The resulting spin-torque effect on the free layer will be either of "anti-damping" or "extra-damping" type, experimentally observable as an additional increase/decrease of the linewidth in the antiparallel/parallel orientation. It is worth noting that the acoustic regime is in a full analogy to the case of a magnetically uncoupled FM1/NM/FM2 system [5] when the resonances coincide. Another important point is that diffusive regime of the spin transport will suppress the above described effects due to averaging of transversal components of the spin currents.

Our study has also shown that the hybridization effect on the linewidth is of the same magnitude as the non-local damping effect in the case of weak interlayer coupling, and that the hybridization fully dominates in the case of strongly coupled magnetic layers. A separation of these two contributions, however, is possible due to their different angular behavior. In general case, contribution of the hybridization to the linewidth parameter will be dependent on degree of asymmetry of layers. Thus, one can expect that, if the free and fixed layers would have the same damping, the influence of the hybridization would be considerably suppressed.

**Acknowledgements**

This work was partially supported by the FCT of Portugal through the projects PEst/CTM/LA0025/2011, RECI/FIS-NAN/0183/2012, PTDC/CTM-NAN/112672/2009, PTDC/FIS/120055/2010, and grants SFRH/BPD/74086/2010 (A.A.T.) and IF/00981/2013 (G.N.K.) as well as by the European FP7 project "Mold-Nanonet".